\documentclass[a4paper,10pt,twoside]{cpc-hepnp}

\usepackage{multicol}
\usepackage{graphicx}
\usepackage{booktabs}
\usepackage{amssymb,bm,mathrsfs,bbm,amscd}
\usepackage[tbtags]{amsmath}
\usepackage{lastpage}

\begin{document}

\fancyhead[c]{\small Submitted to 'Chinese Physics C'} 
\fancyfoot[C]{\small 010201-\thepage}

\footnotetext[0]{Received 14 March 2009}

\title{Simulation of Cone Beam CT System Based on Monte Carlo Method\thanks{ Supported by National Natural Science Foundation of China (81101132, 11305203) and Natural Science Foundation of Anhui Province(11040606Q55, 1308085QH138) }}

\author{%
       \email{liqin.hu@fds.org.cn}%
\quad WANG Yu(ÍõÓñ)$^{1,2)}$
\quad CHEN Chaobin(³Â³¯±ó)$^{2}$
\quad CAO Ruifen(²ÜÈð·Ò)$^{2}$\\
\quad HU Liqin(ºúÀöÇÙ)$^{1,2;1)}$
\quad LI Bingbing(Àî±ø±ø)$^{3}$
}
\maketitle

\address{%
$^1$ University of Science and Technology of China, Hefei, 230027, China\\
$^2$ Institute of Nuclear Energy Safety Technology, Chinese Academy of Sciences, Hefei, 230031, China\\
$^3$ Cancer Hospital, HeFei Institutes of Physical Science, Chinese Academy of Science, Hefei, 230031, China\\
}

\begin{abstract}
Adaptive Radiation Therapy (ART) was developed based on Image-guided Radiation Therapy (IGRT) and it is the trend of photon radiation therapy. To get a better use of Cone Beam CT (CBCT) images for ART, the CBCT system model was simulate established based on Monte Carlo program and validated against the measurement. The BEAMnrc program was adopted to the KV x-ray tube. Both IOURCE-13 and ISOURCE-24 were chosen to simulate the path of beam particles. The measured Percentage Depth Dose (PDD) and lateral dose profiles under 1cm water were compared with the dose calculated by DOSXYZnrc program. The calculated PDD was better than 1\% within the depth of 10cm. The errors of more than 85\% points of lateral dose profiles were within 2\%. The CBCT system model helps to improve CBCT image quality for dose verification in ART and assess the CBCT image concomitant dose risk.
\end{abstract}



\footnotetext[0]{\hspace*{-3mm}\raisebox{0.3ex}{$\scriptstyle\copyright$}2013
Chinese Physical Society and the Institute of High Energy Physics
of the Chinese Academy of Sciences and the Institute
of Modern Physics of the Chinese Academy of Sciences and IOP Publishing Ltd}%

\begin{multicols}{2}

\section{Introduction}

Image-guided Radiation Therapy (IGRT) technique had improved the accuracy of radiation therapy significantly[1], which primarily used Cone Beam Computed Tomography (CBCT) to monitor and track the patient positioning and the target shape in clinical practice. Adaptive Radiation Therapy (ART) was developed based on IGRT and was the trend of photon radiotherapy. The ideal ART was to monitor and subsequently correct each variables in the radiation therapy  process based on real time image data, for example, the CBCT images[2]. However, a large number of artifacts contained in CBCT images obscure the tissue information and further limit CBCT images for ART. On the other hand, the risk assessment of CBCT concomitant dose has also become a global concern[3].

It is widely accepted that Monte Carlo is the most accurate tool to estimate the dose distribution in the phantom currently. A number of research units constantly tried to use Monte Carlo program to establish CBCT system model in past 10 years[4-6]. The right CBCT system model would be useful to improve the CBCT image quality and make an accurate CBCT concomitant dose risk assessment.

FDS Team, an interdisciplinary research team, has been committed to advanced nuclear energy and physics techniques, mainly including nuclear reactor physics[7-9], nuclear reactor material[10,11], nuclear reactor engineering[12,13], numerical simulation and visualization[14], medical physics and environment protection etc.

FDS Team has focused on the research of accurate/advanced physics technologies of radiotherapy for over 10 years and has developed a series of accurate radiotherapy planning system, named Accurate/Advanced Radiotherapy System (ARTS)[15], involving the key technologies including multi-objective optimization[16,17], fast Monte Carlo dose calculation[13], dynamic target tracking[18], 3D dose reconstruction[19] and digital modeling[20-22]. The experimental platform consisted of electron linear accelerator, 2D ionization chamber and 3D water tank for dose verification equipment, laser and infrared positioning and tracking system, anthropomorphic phantom and so on.

The aim of this work was to simulate the clinical CBCT system based on Monte Carlo method and validate the model based on ARTS experimental platform.

\section{Material and methods}

\subsection{ELEKTA AXESSE XVI CBCT system}

In this work we present the Monte Carlo simulation of XVI CBCT system equipped with ELEKTA AXESSE electron linear accelerator (ELEKTA, Sweden). The XVI system consists of a KV x-ray tube and KV flat panel detector. The collimator and filter combination could form different field of view (FOV), which shaped different image size in the detector. The image size could be classified to be small FOV, medium FOV and large FOV according to collimator size. The  image center of small FOV was in the beam central axis, while the image center of large FOV and medium FOV was deviated from beam central axis.

There were lots of documents about Monte Carlo simulation of KV x-ray detector[23-25] and the  detector simulation was relatively simple. However, the right x-ray tube model determined the feasibility and reliability of the CBCT system model. For this reason this work was mainly about the simulation of x-ray tube of XVI system and water phantom test was used to validate the model.

\subsection{Monte Carlo simulation of x-ray tube}

Since there was a ${3.5^{\circ}}$ oblique angle of x-tube housing, which would change the particles path and produce an unnegligible effect on dose calculation, so the x-ray tube simulation was divided into two parts. The first part was tube housing simulation, including KV source, target and inherent cone filter. The second part was simulation of collimator, filter, and plastic cover.
\begin{center}
\includegraphics[width=8.5cm]{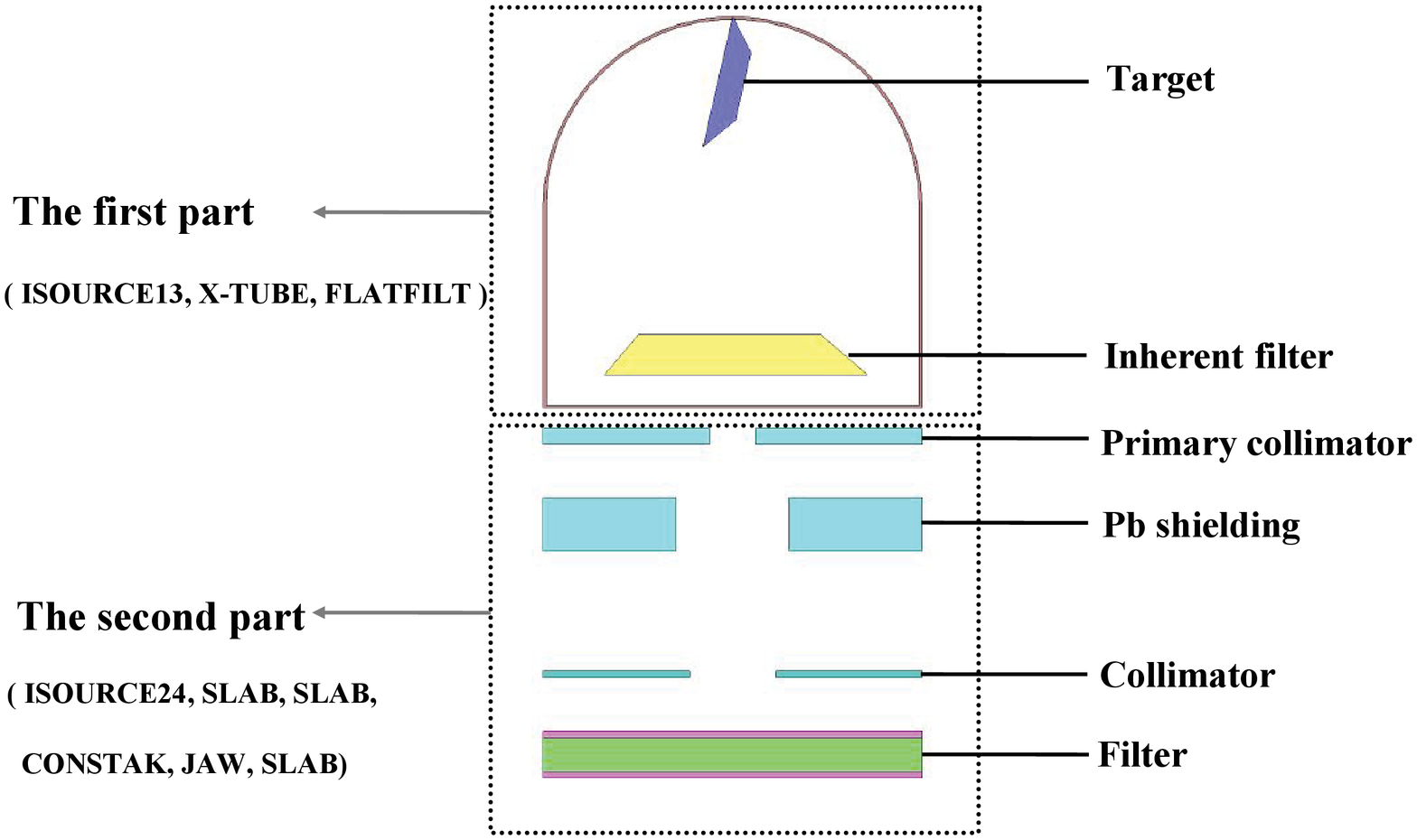}
\figcaption{\label{Fig.1.}   The schematic of X-ray tube model for XVI. }
\end{center}

\begin{center}
\tabcaption{ \label{tab1} The CMs and material used for X-ray tube model.}
\footnotesize
\begin{tabular*}{85mm}{l@{\extracolsep{\fill}}lll}
\toprule \bf{Geometry}   & \bf{CMs for simulation}   & \bf{Material}   \\
\hline
Target             & X-TUBE   & W/Graphite  \\
Inherent filter     & FLATFILT & Al/Al/Al/Cu \\
Primary collimator  & CONSTAK  & Pb          \\
Collimator          & JAW      & Pb          \\
Filter              & SLAB     & PETG/Air/PETG \\
Plastic cover       & SLAB     & Polystyrene \\
\bottomrule
\end{tabular*}
\end{center}
\normalsize

The EGS subroutine BEAMnrc program provided different component modules (CMs)[27] for simulating different geometry structures. The target was generally simulated by X-TUBE CM. The collimators could be simulated by any of BLOCK CM, JAW CM, CONSTAK CM and MLC CM. The conical flattening filter was simulated using FLATFILT CM.The multilayer plate perpendicular to Z axis was simulated by SLABS CM. The model of x-ray tube of XVI system was shown in Fig.1 and the component modules and material adopted for simulation were detailed in Tab.1.

To start the simulation, the reaction cross section data of different material with Rayleigh scattering information should be generated using EGS subroutine PEGS4 program[28].


\subsubsection{Monte Carlo simulation of the first part}

According to the Corrective Maintenance Manual for XVI4.5 [29], the first part of x-ray tube included KV source, target and inherent filter (see Fig.1). The incident x-ray source was simulated using ISOURCE-13 rectangular beam with a monochromatic energy of 120KV. The target had an angle of ${14^{\circ}}$ and the length and width of focus spot were 0.04cm and 0.08cm respectively. Inherent filter was simplified with 10 layers, including exit window layer, Aluminum layer, Vacuum layer and so on.

To improve the simulation efficiency and guarantee the effective particles number for subsequent calculation, the default parameters in the macro file beamnrc\underline{\hbox to 1.0mm{}} user\underline{\hbox to 1.0mm{}}macros.mortran were modified.
The value of boundary tolerance BDY\underline{\hbox to 1.0mm{}}TOL was modified to ${5\times10^{-7}}$ and the maximum directional bremsstrahlung split number be ${2\times10^{4}}$ according to the suggestion of Ali and Rogers [30]. Some Variance Reduction Techniques were adopted, such as Rayleigh scattering, Atomic relaxations, Bound Compton scattering, Electron impact ionization and Spin effects. The global Electron Cutoff energy (ECUT) and Photon Cutoff energy (PCUT) set to 0.521MeV and 0.01MeV.

\subsubsection{Monte Carlo simulation of the second part}

The second part of X-ray tube mainly included primary collimator, collimator, filter and plastic cover. ISOURCE-24 was chosen to simulate the oblique phase space source obtained from the first part simulation. CONSTAK CM was used to simulate the primary collimator, which had 3 layers. The collimator was simulated using JAW CM, which had an air layer of ${5.5\times5.5cm^{2}}$ in the middle. The filter was simulated by SLABS CM with 5 layers. The detailed information for each layer of CMs was shown in Fig.1. Different CMs could be used to simulate the same geometry, for example, any of BLOCK, JAW and CONSTAK CM could be used to simulate collimator, but different choice had no significant effect on the calculation results.

\subsubsection{Water phantom simulation}

The dose distribution in a water phantom for XVI model was calculated by EGS subroutine DOSXYZnrc program, the grid size set to ${0.5\times0.5\times0.2cm^{3}}$ for dose calculation. The option of HOWFARLESS was selected and photon split number N\underline{\hbox to 1.0mm{}}split was 100, the history set to ${100\times10^{6}}$.

\subsection{Data measurement}

The correctness of x-ray tube model was verified against measured dose for percentage depth dose and lateral profile at certain depth. For the measurements the x-ray tube was at a standstill of ${270^{\circ}}$, which was the accelerator gantry position.The 120KV beam was delivered for 300 frames, a nominal 40mA and a nominal 40ms per frame. The radiation preset parameter was just as Tab.2.
\begin{center}
\tabcaption{ \label{tab2}  The preset parameters for XVI radiation.}
\footnotesize
\begin{tabular*}{80mm}{l@{\extracolsep{\fill}}lll}
\toprule Radiation Preset   & Parameter value   \\
\hline
Image acquisition type    & PlanarView  \\
Energy                    & 120KV \\
Total Frames              & 300          \\
NominalmAPerFrame         & 40mA          \\
NominalmsPerFrame         & 40ms \\
Collimator                & S20 \\
Filter                    & F0 \\
\bottomrule
\end{tabular*}
\end{center}

After calibration, 3D Blue Water Phantom (IBA, Germany) was put right under the x-ray tube. The source-to-surface distance (SSD) was at 75cm. When the radiation was delivered, the dose was measured from depth 0cm to 20cm. 10 times measurement for each point and then average value  stored. Then with the same radiation condition the planar dose under 1cm RW3 solid water (PTW, Germany) was measured using PTW729 ionization chamber (PTW, Germany).

\section{Results }

The simulation of x-ray tube was validated by two aspects. (1) EGS subroutine BEAMdp program was adopted to calculate the spectrum distribution and energy fluence, the analysis results of the model were compared with existing literatures, (2) dose measurement of water phantom.

\subsection{BEAMdp analysis of x-ray tube model}
As seen in Fig.1, the phase space file obtained from the second part simulation was analyzed by BEAMdp program. The spectrum distribution and energy fluence were shown in Fig.2 and Fig.3.

When high speed electron impinged on the target material, the characteristic and bremsstrahlung radiation were produced. There would be a larger proportion of characteristic x-ray superimposed on a continuous x-ray lines in spectrum distribution. The characteristic x-ray could tell the information of the target. Generally the characteristic x-ray of tungsten was 59.3KeV, accounted for 57.97\%[31].

As seen in Fig.2, the energy peaked at about 60KeV, which was consistence with the characteristic x-ray of tungsten target.
\begin{center}
\includegraphics[width=6.0cm]{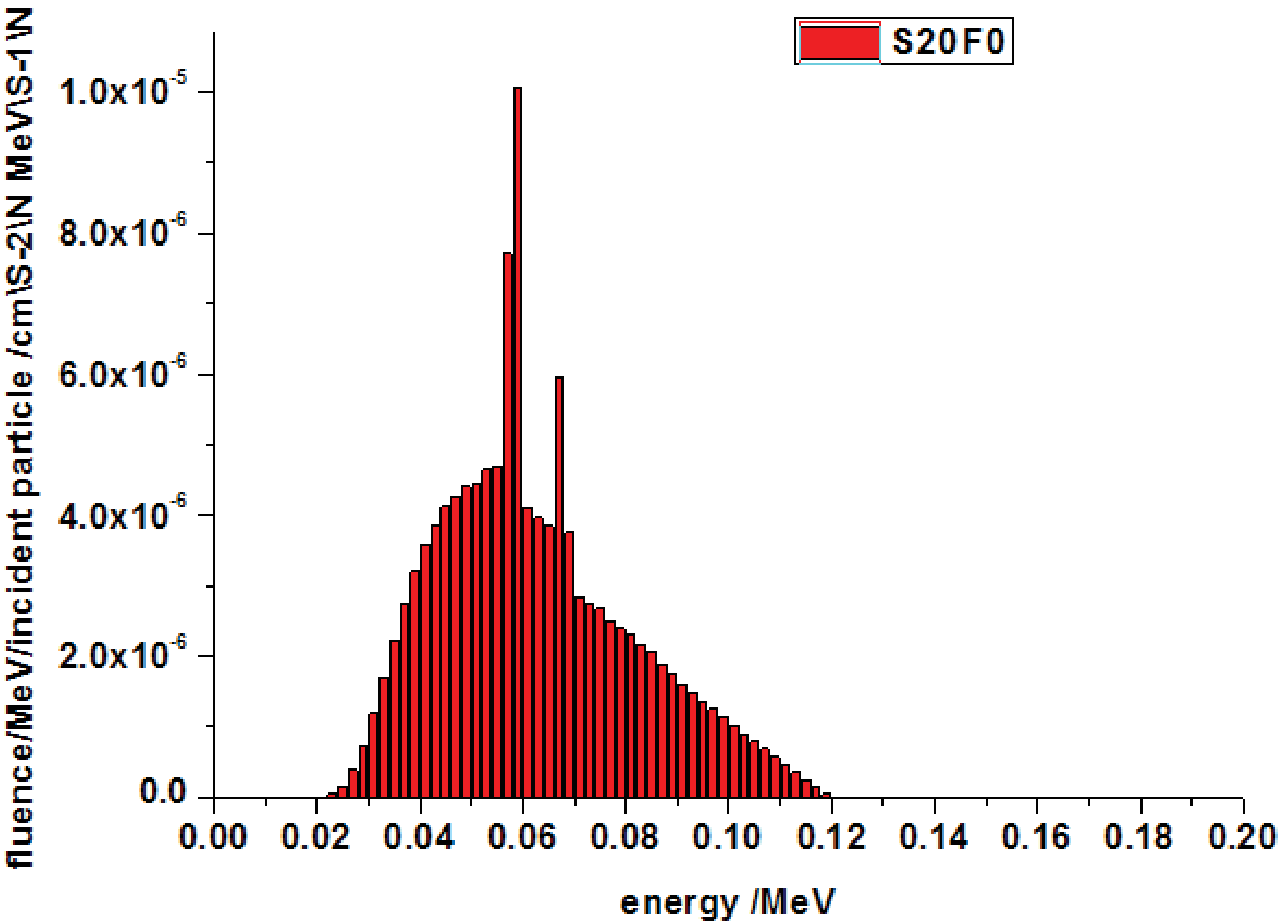}
\figcaption{\label{Fig.2.} The spectrum distribution for XVI with 120KV. }
\end{center}

Fig.3 illustrates the energy fluence profile for S20F0. All these results were in accordance with the results of E.Spezi[32].
\begin{center}
\includegraphics[width=6.0cm]{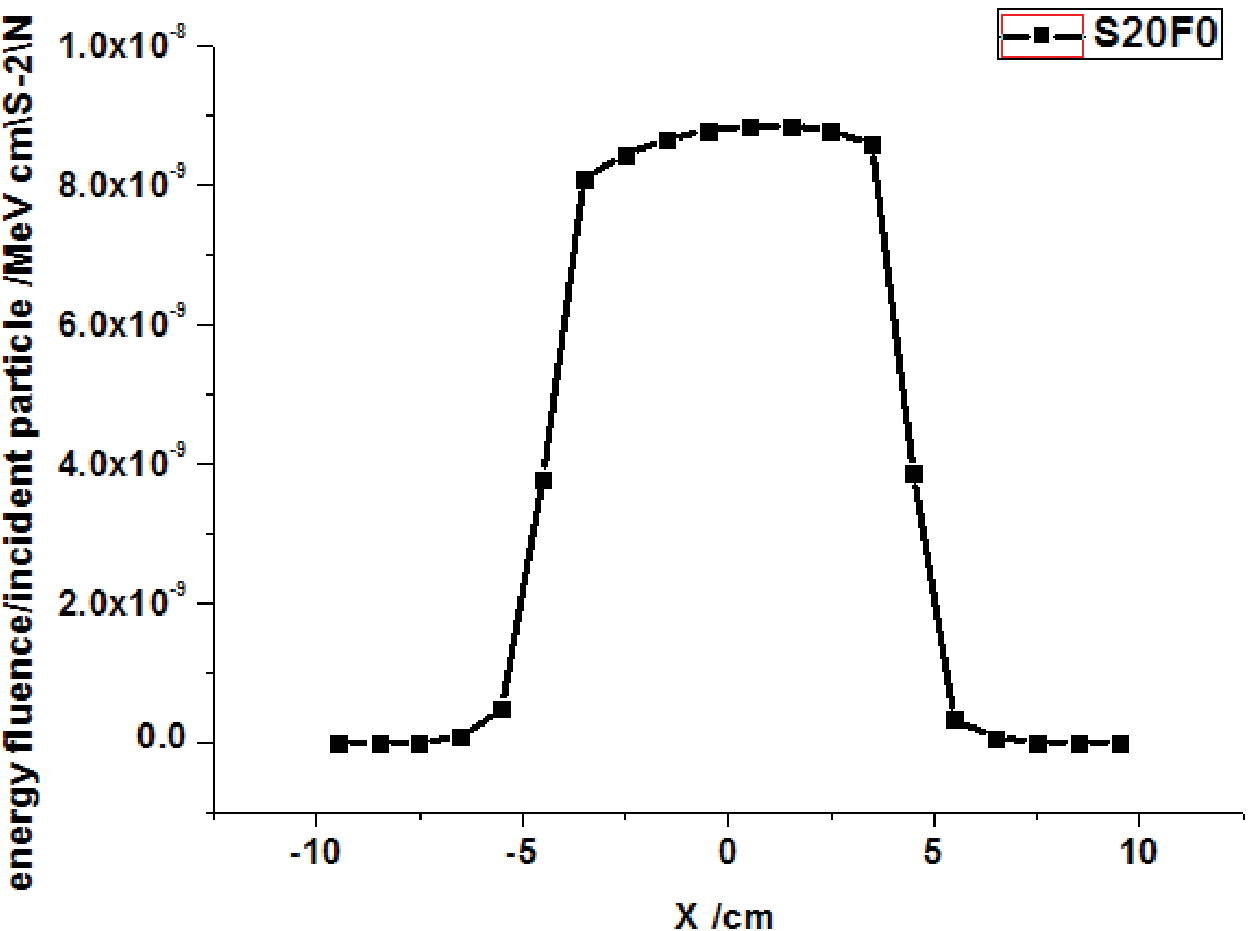}
\figcaption{\label{Fig.3.} The spectrum distribution for XVI with 120KV. }
\end{center}
\subsection{Measurement vs Calculation}
The measured lateral dose data obtained from PTW729 ionization chamber was a ${27\times27}$ matrix with a spacing of 1cm. The calculated data was ${60\times60\times100}$, the spacing of adjacent point in three axis was 0.5cm, 0.5cm and 0.2cm respectively. All the data was normalized by the maximum dose.

Fig.4 and Fig.5 depicts that the error was less than 1\% in intervals of [-15cm, 10cm] and [10cm, 15cm] in cross-line direction. The error was less than 2\% in intervals of [-15cm, 10cm], [-7cm,7cm] and[10cm,15cm] in in-line direction. Note that a slightly bigger error appeared around points of -10cm and 10cm in off-axis direction. It would be caused by two main reasons. Firstly, the data acquisition software of PTW729 ionization chamber possessed the function of automatic smooth and uniform data processing, which introduced the measurement error. Secondly, there was a protective layer about 3mm thickness on the surface of PTW729 ionization chamber. The data measured at 1cm depth was actually at depth of 1.3cm while the data calculated was at 1cm depth.
\begin{center}
\includegraphics[width=5.0cm]{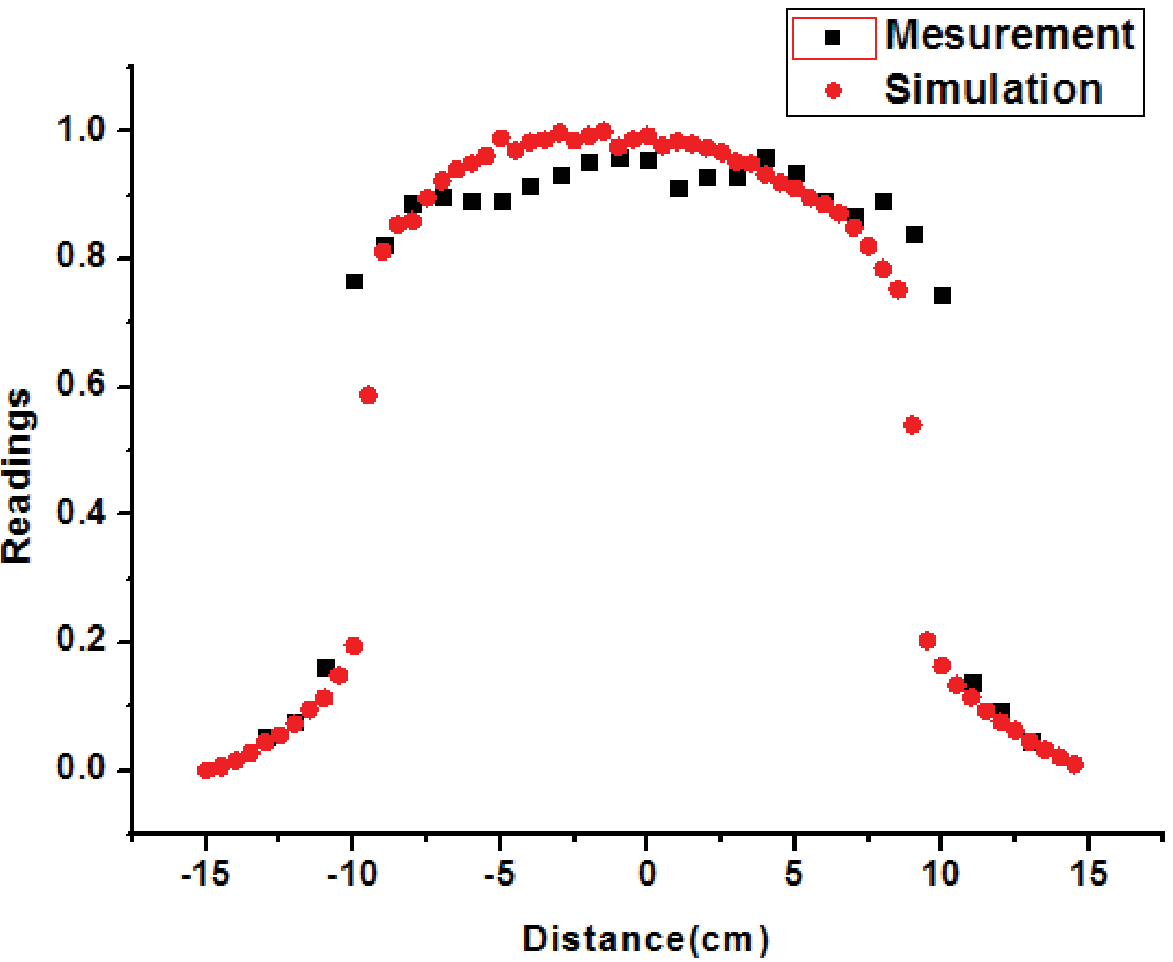}
\figcaption{\label{Fig.4.} The dose comparison in cross-line direction under 1cm water. }
\end{center}
\begin{center}
\includegraphics[width=5.0cm]{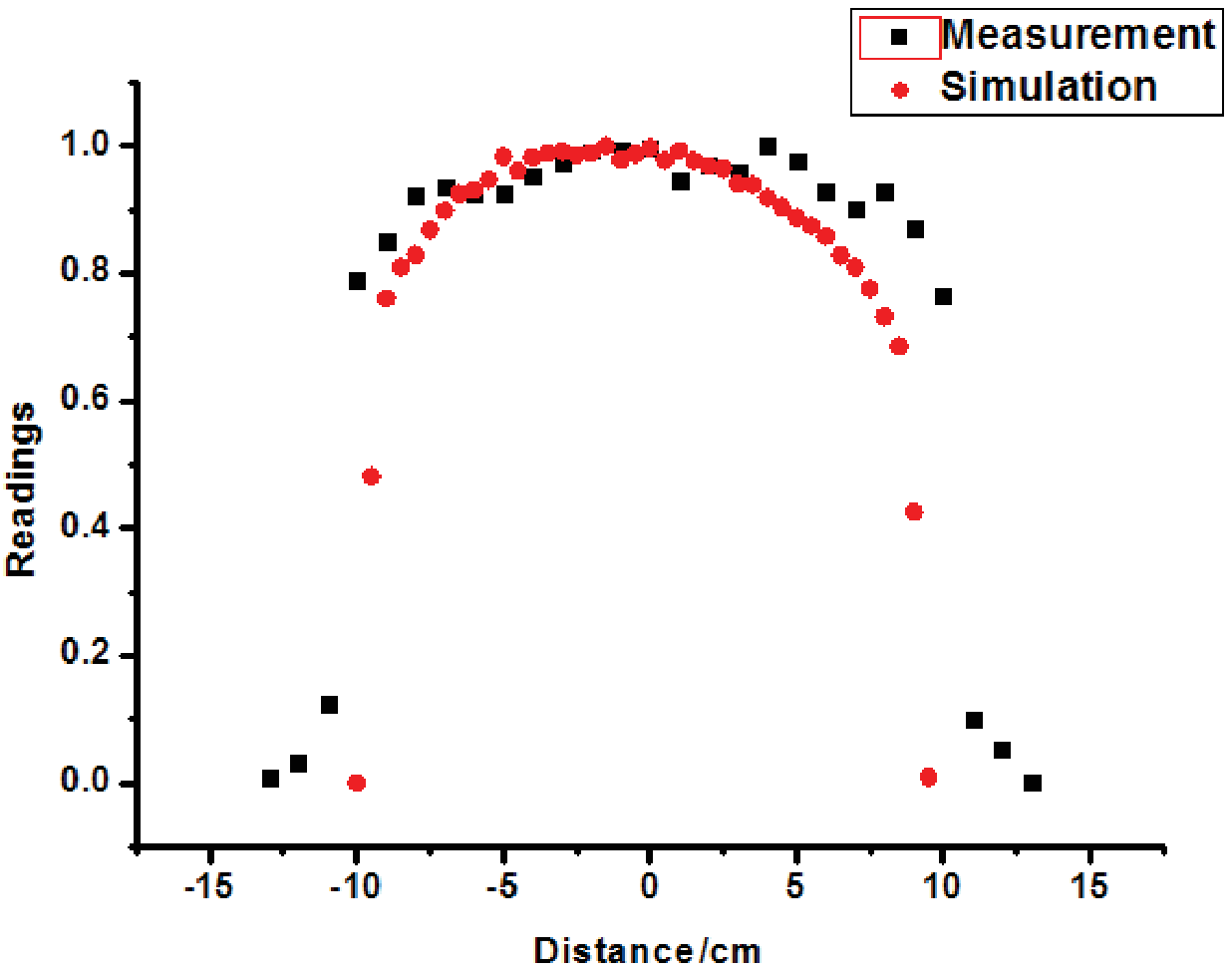}
\figcaption{\label{Fig.5.} The dose comparison in in-line direction under 1cm water. }
\end{center}

Fig.6 shows that the PDD calculation was in a good agreement with measurement. The error was less than 1\% within 10cm thickness. However, as the phantom thickness increased beyond 10cm the error increased, the error was up to 4\% for 20cm, that was because multiple scatter existed in phantom radiation but it was not taken into consideration in calculation.
\begin{center}
\includegraphics[width=5.0cm]{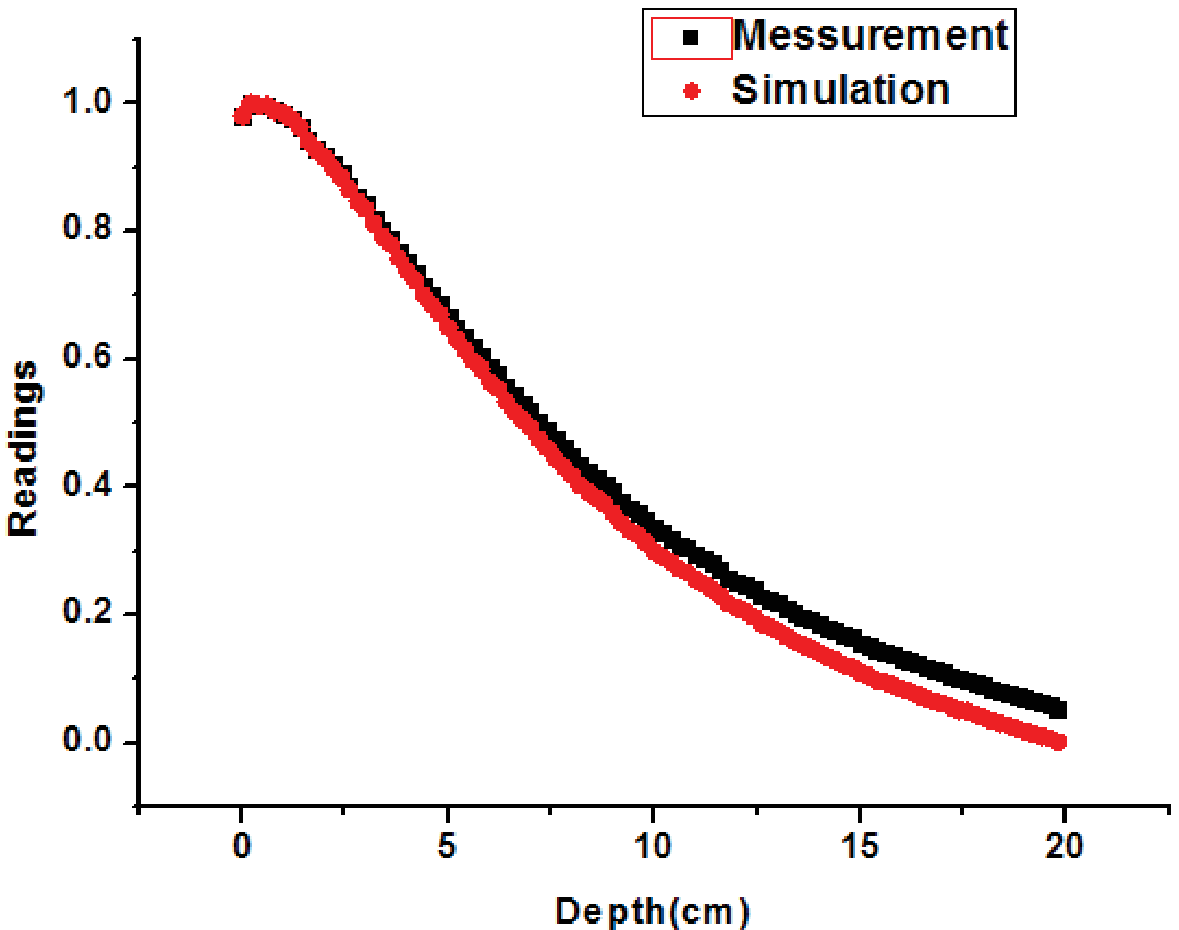}
\figcaption{\label{Fig.6.} The comparison of percentage depth dose for XVI radiation. }
\end{center}

Fig.7 shows XZ planar dose distribution of calculation. There was almost no dose deposition beyond 15cm thickness along the beam central axis. The error distribution of calculation was seen in Fig.8, it was obvious that the error were less than 1\% in [-8cm, 8cm] interval along Z axis in cross-line direction. Beyond the depth of 10cm the error becomes bigger, even up to 3\%.
\begin{center}
\includegraphics[height=5.0cm]{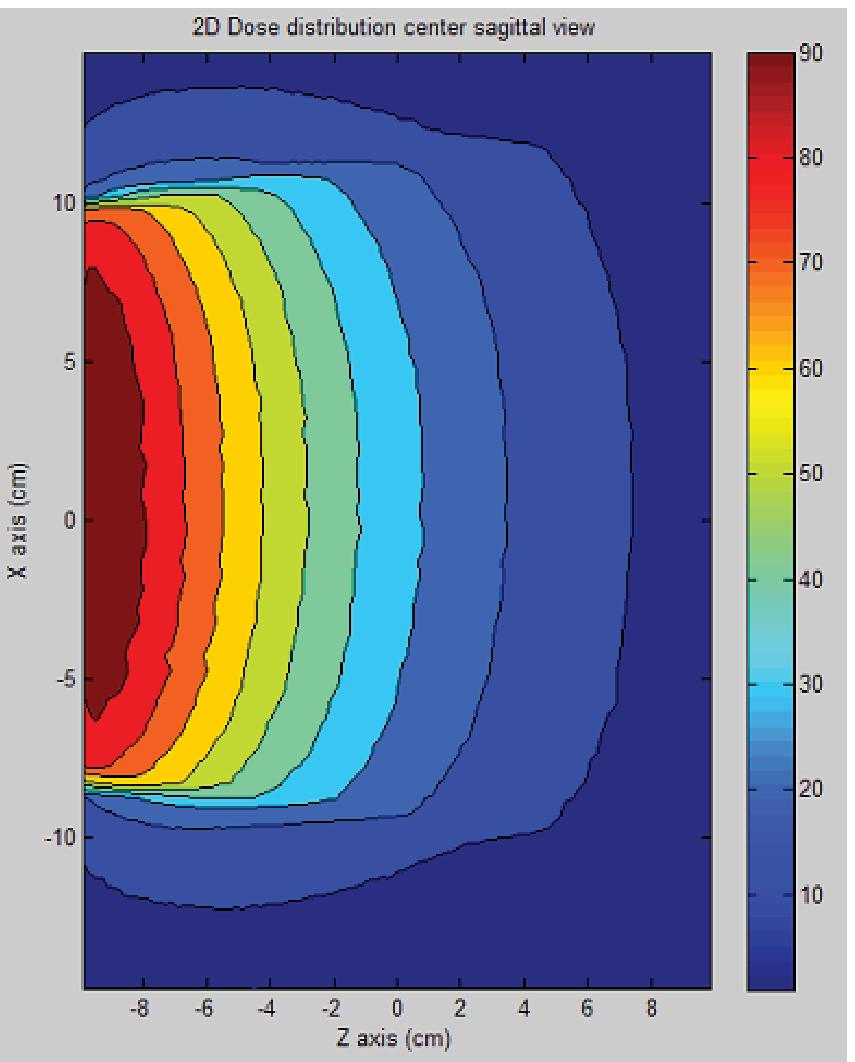}
\figcaption{\label{Fig.7.} The calculation XZ planar dose profile. }
\end{center}
\begin{center}
\includegraphics[height=5.0cm]{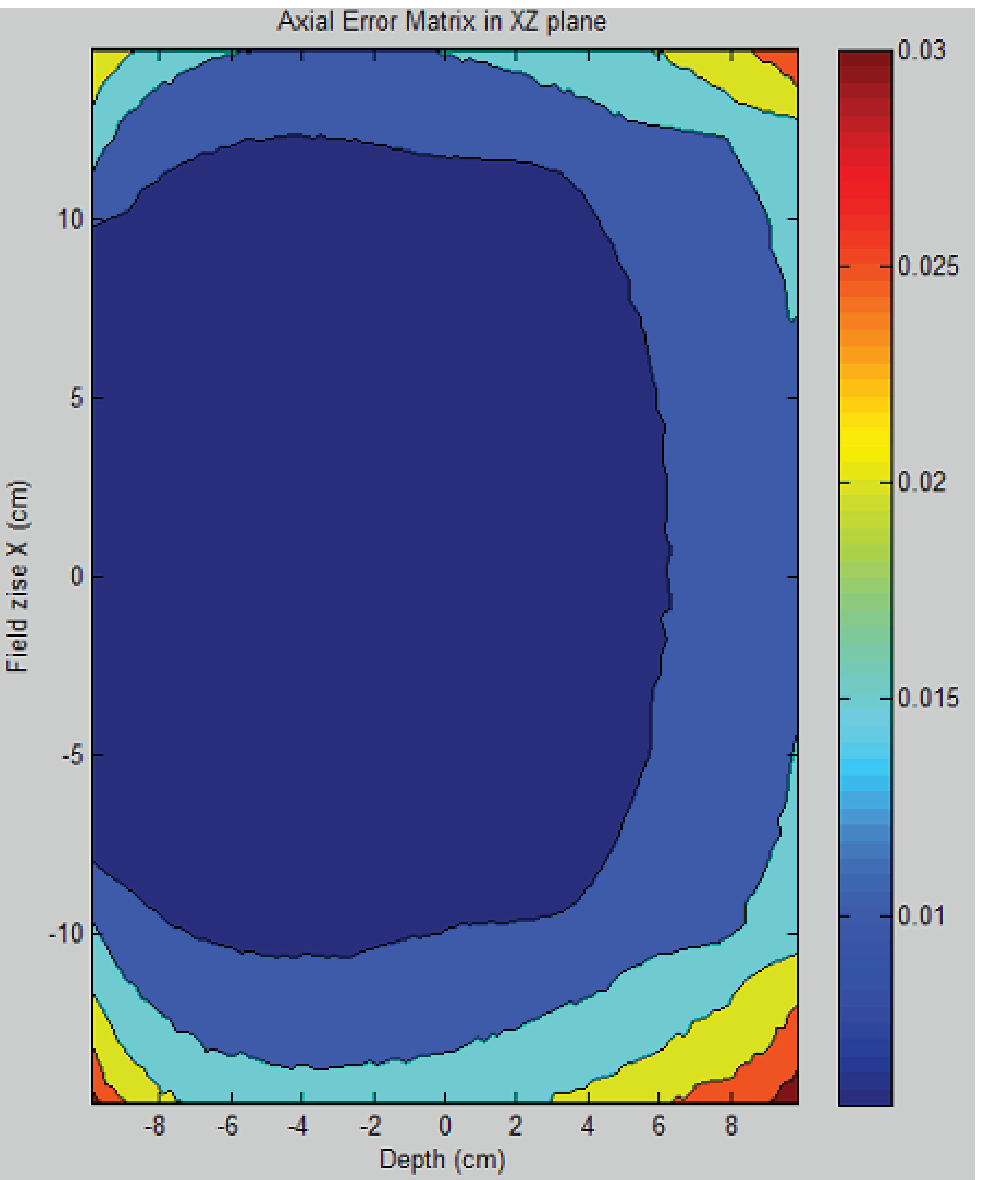}
\figcaption{\label{Fig.8.} The error profile for XZ planar dose calculation. }
\end{center}

\section{Discussion and Conclusion }

The XVI CBCT system model was established and the simulation was validated against measurements based on ARTS experimental platform.

The PDD accuracy within 10cm was better than 1\%. More than 85\% lateral points have a good conformity with measurement. Note that the error was significant around the points of -10cm and 10cm in lateral dose profile, which was caused by the automatic data processing of the software and actually measuring a deeper position than calculation for there was a protective layer on the chamber surface. The error could be eliminated by optimizing the simulation parameters and improving measurement accuracy, such as increasing the effective particles number, increasing measure times and closing automatic data processing function of the software.

The CBCT system model would make CBCT image quality improvement and further adaptive radiotherapy and CBCT concomitant dose risk assessment possible. In further work, the different image acquisition modes, which depended on the variables of voltage, current, frames, phantom thickness, KV detector response and so on, would be investigated to get a better known of clinical CBCT system.

\section{Acknowledgments}

This paper was supported by National Natural Science Foundation of China (81101132 and 11305203) and Natural Science Foundation of Anhui Province(11040606Q55 and 1308085QH138).

The author wishes to thank FDS Team members for their help in this study.

\end{multicols}

\clearpage

\end{document}